\documentclass[11pt,a4paper]{article}

\usepackage{moriond,epsfig}

\usepackage{xspace}

\usepackage{caption}
\usepackage{subfig}

\usepackage{ifthen}
\newboolean{uprightparticles}
\setboolean{uprightparticles}{true}
\usepackage{amssymb}
\usepackage{amsfonts}
\usepackage{upgreek}

\def\lhcb  {LHCb\xspace}
\def\LHCb  {\lhcb}
\def\cern  {CERN\xspace}
\def\lhc   {LHC\xspace}

\ifthenelse{\boolean{uprightparticles}}%
{
 \def\Ppi      {\ensuremath{\uppi}\xspace}

 \def\PB      {\ensuremath{\mathrm{B}}\xspace}                 
 \def\PD      {\ensuremath{\mathrm{D}}\xspace}                 
 \def\PK      {\ensuremath{\mathrm{K}}\xspace}

 \def\Pp      {\ensuremath{\mathrm{p}}\xspace}                 
}
{
 \def\Ppi      {\ensuremath{\pi}\xspace}

 \def\PB      {\ensuremath{B}\xspace}                 
 \def\PD      {\ensuremath{D}\xspace}                 
 \def\PK      {\ensuremath{K}\xspace}

 \def\Pp      {\ensuremath{p}\xspace}                 
}

\def\pion  {\ensuremath{\Ppi}\xspace}
\def\pip   {\ensuremath{\pion^+}\xspace}
\def\pim   {\ensuremath{\pion^-}\xspace}

\newcommand{\pislow}{\ensuremath{\pion_{\mathrm{slow}}}\xspace}

\def\kaon  {\ensuremath{\PK}\xspace}
\def\Kp    {\ensuremath{\kaon^+}\xspace}
\def\Km    {\ensuremath{\kaon^-}\xspace}

\def\Dbar    {\kern 0.2em\overline{\kern -0.2em \PD}{}\xspace}
\def\D       {\ensuremath{\PD}\xspace}

\def\Dz      {\ensuremath{\D^0}\xspace}
\def\Dzb     {\ensuremath{\Dbar^0}\xspace}
\def\DzDzb   {\ensuremath{\Dz {\kern -0.16em \Dzb}}\xspace}
\def\Dp      {\ensuremath{\D^+}\xspace}
\def\Dm      {\ensuremath{\D^-}\xspace}

\def\DpDm    {\ensuremath{\Dp {\kern -0.16em \Dm}}\xspace}
\def\Dstar   {\ensuremath{\D^*}\xspace}

\def\Dstarp  {\ensuremath{\D^{*+}}\xspace}
\def\Dstarm  {\ensuremath{\D^{*-}}\xspace}

\def\Dpors      {\ensuremath{\D^+_{(s)}}\xspace}

\def\B       {\ensuremath{\PB}\xspace}
\def\Bbar    {\kern 0.18em\overline{\kern -0.18em \PB}{}\xspace}

\def\proton      {\ensuremath{\Pp}\xspace}

\def\CP                {\ensuremath{C\!P}\xspace}

\def\CPV               {\ensuremath{C\!PV}\xspace}

\newcommand{\ARaw}{\ensuremath{{\cal A}_{\mathrm{Raw}}}\xspace}
\newcommand{\ARawstar}{\ensuremath{\ARaw^{\ast}}\xspace}
\newcommand{\AD}{\ensuremath{{\cal A}_{\mathrm{D}}}\xspace}
\newcommand{\AP}{\ensuremath{{\cal A}_{\mathrm{P}}}\xspace}

\newcommand{\Delm}{\mbox{$\Delta m $}\xspace}
\newcommand{\ACP}{\ensuremath{{\cal A}_{\CP}}\xspace}

\def\ycp        {\ensuremath{y_{\CP}}\xspace}
\def\yCP        {\ycp}
\def\agamma     {\ensuremath{A_{\Gamma}}\xspace}
\def\AGamma     {\agamma}

\newcommand{\tev}{\ensuremath{\mathrm{\,Te\kern -0.1em V}}\xspace}
\newcommand{\gev}{\ensuremath{\mathrm{\,Ge\kern -0.1em V}}\xspace}
\newcommand{\mev}{\ensuremath{\mathrm{\,Me\kern -0.1em V}}\xspace}
\newcommand{\kev}{\ensuremath{\mathrm{\,ke\kern -0.1em V}}\xspace}
\newcommand{\ev}{\ensuremath{\mathrm{\,e\kern -0.1em V}}\xspace}
\newcommand{\gevc}{\ensuremath{{\mathrm{\,Ge\kern -0.1em V\!/}c}}\xspace}
\newcommand{\mevc}{\ensuremath{{\mathrm{\,Me\kern -0.1em V\!/}c}}\xspace}
\newcommand{\gevcc}{\ensuremath{{\mathrm{\,Ge\kern -0.1em V\!/}c^2}}\xspace}
\newcommand{\gevgevcccc}{\ensuremath{{\mathrm{\,Ge\kern -0.1em V^2\!/}c^4}}\xspace}
\newcommand{\mevcc}{\ensuremath{{\mathrm{\,Me\kern -0.1em V\!/}c^2}}\xspace}

\def\mbarn{\ensuremath{\rm \,mb}\xspace}

\def\invpb {\ensuremath{\mbox{\,pb}^{-1}}\xspace}

\def\invfb   {\ensuremath{\mbox{\,fb}^{-1}}\xspace}

\def\lumiyr {\ensuremath{37.7\,\invpb}\xspace}
\def\lumiacp {\ensuremath{37\,\invpb}\xspace}

\begin{document}

\markboth{Patrick Spradlin}
{Prospect of \Dz mixing and \CPV at \LHCb}

%
%

\title{Prospect of \Dz mixing and \CPV at \LHCb}

\author{Patrick Spradlin\footnote{
On behalf of the \LHCb collaboration.}}

\address{Particle Physics Department, University of Oxford, Denys Wilkinson Building, Keble Road\\
Oxford OX1 3RH,
United Kingdom\\
Patrick.Spradlin@cern.ch}

\maketitle


\begin{abstract}
  Precision measurements in charm physics offer a window into a unique 
  sector of potential New Physics interactions.
  \LHCb is poised to become a world leading experiment for charm studies,
  recording enormous statistics with a detector tailored for flavor physics.
  This article presents recent charm \CPV and mixing studies from \LHCb,
  including \LHCb's first \CP asymmetry measurement with \lumiacp
  of data collected in 2010.
  The difference of the \CP asymmetries of \Dz decays
  to the $\Km \Kp$ and $\pim \pip$ final states is determined to be
  $\Delta\ACP = \left(-0.28 \pm 0.70 \pm 0.25 \right)\%$.
  Significant updates to the material presented at the 4$^{\rm th}$
  International Workshop on Charm Physics are included.
\end{abstract}


\section{The \LHCb experiment}
\label{sec:LHCb}

  \LHCb, the dedicated flavor experiment at \cern's Large Hadron Collider
  (\lhc), is the only \lhc experiment currently performing measurements of
  charm \CP violation (\CPV) and $\Dz$-$\Dzb$ mixing.
  Many of the features that make \LHCb an excellent \B-physics laboratory also
  make it well-suited for precision charm physics studies.\cite{Alves:2008zz}
  The cross-section to produce charm hadrons into the \LHCb acceptance
  in the LHC's $\sqrt{s} = 7\,\tev$ proton-proton collisions is
  \mbox{$1.23 \pm 0.19\,\mbarn$}, creating a huge potential data
  set.\cite{LHCb-CONF-2010-013}
  The \LHCb trigger system has a flexible design that includes
  dedicated charm triggers so that this prolific production can be exploited.

  \LHCb recorded a total integrated luminosity of \lumiyr in 2010.
  These data were collected under rapidly evolving interaction conditions as
  the \lhc provided high quality beams with increasing bunch numbers and
  intensities.
  Large charm data sets were collected in 2010, and the 2011-12 run promises
  to yield even larger samples of charm decays, with a target of $1\,\invfb$.

\section{Time-integrated \CPV in \D mesons}
\label{sec:cpv}

  \LHCb is searching for evidence of new sources of \CP asymmetry in the
  time-integrated decay rates of \D mesons.
  For a given final state $f$, the time-integrated \CP asymmetry, $\ACP(f)$,
  is defined as
  \begin{equation}
    \ACP(f) = \frac{\Gamma(\D \to f) - \Gamma(\Dbar \to \bar{f})}{\Gamma(\D \to f) + \Gamma(\Dbar \to \bar{f})}.
    \label{eq:cpv:acp}
  \end{equation}
  For \Dpors mesons this is a measurement of direct \CPV , while
  \Dz decays may have contributions from both indirect and direct \CPV.
  In the Standard Model, \CPV in the charm system is highly suppressed.
  Indirect \CPV is negligibly small and should be common for all decay modes.
  Direct \CPV is expected to be $\mathcal{O}(10^{-3})$ or less and to vary
  among decay modes.\cite{Bianco:2003vb}
  In \CPV searches in singly Cabibbo suppressed decays,
  such as $\Dz \to \Km \Kp$, participation of well-motivated new physics
  (NP) particles in the interfering penguin amplitude
  could enhance direct \CPV up to $\mathcal{O}(10^{-2})$.\cite{Grossman:2006jg}

  \LHCb recently presented its first results of time-integrated \CPV
  measurements in decays $\Dz \rightarrow \Km \Kp$ and
  $\Dz \rightarrow \pim \pip.$\cite{LHCb-CONF-2011-023}
  The analysis uses \Dz mesons reconstructed as the product of
  $\Dstarp \rightarrow \Dz \pislow^+$ decays so that their initial flavors are
  identified (tagged) as \Dz or \Dzb by the charge of the tagging slow pion.
  The asymmetries of the \Dstar-tagged raw yields, $\ARawstar$, can be written
  as a sum of components:
  \begin{equation}
    \ARawstar(f)  =  \ACP(f) + \AD(f) + \AD(\pislow) + \AP(\Dstarp),\label{eq:cpv:comp:tag}
  \end{equation}
  where $\AD(f)$ and $\AD(\pislow)$ are the detection asymmetries of the final
  state $f$ and the tagging pion $\pislow^{\pm}$ respectively and
  $\AP(\Dstarp)$ is the production asymmetry of \Dstarp.
  For the self-conjugate final states $\Km\Kp$ and $\pim\pip$,
  $\AD(\Km\Kp) = \AD(\pim\pip) = 0$.
  The production asymmetries \AP are independent of final state, as is 
  $\AD(\pislow)$.
  Hence, the difference in $\ACP(f)$ for $f = \Km\Kp$ and $\pim\pip$ can
  be measured precisely with the confounding systematic asymmetries
  canceling exactly:
  \begin{eqnarray}
    \Delta\ACP   & \equiv & \ACP(\Km\Kp) - \ACP(\pim\pip),\label{eq:cpv:delacp:def} \\
                 & = & \ARawstar(\Km\Kp) - \ARawstar(\pim\pip).\label{eq:cpv:delacp:raw}
  \end{eqnarray}
  In \lumiacp of \LHCb 2010 data, we measure $\Delta\ACP$ consistent with zero:
  \begin{equation}
    \Delta\ACP = \left(-0.28 \pm 0.70 \pm 0.25 \right)\%,\label{eq:cpv:result}
  \end{equation}
  where the first uncertainty is statistical and the second is systematic.
  This result is approaching the sensitivity of \CPV measurements performed by
  the \B-factories in these decay modes,\cite{Aubert:2007if,Staric:2008rx}
  but not yet at the level of CDF's recent measurement.\cite{CDF-10296}
  Due to differential proper-time acceptance between the $\Km\Kp$ and
  $\pim\pip$ samples, the measured value of $\Delta\ACP$ includes a residual
  $10\%$ of the mode-independent indirect \CP asymmetry.
  No limiting systematic bias has been identified in the method, so
  future iterations of the measurement with the much larger data set
  anticipated for 2011-2012 will be significantly more precise.

\section{Time-dependent \CPV and mixing measurements in \Dz}
\label{sec:mix}

  The conventional parameterization of charm mixing is fully explained
  elsewhere.\cite{Nakamura:2010zzi:D0mix}
  Briefly, the mass eigenstates of the neutral \D system $\D_1$ and $\D_2$
  are expressed as normalized superpositions of the flavor eigenstates \Dz
  and \Dzb:
  \begin{equation}
    \D_1 = p \Dz + q \Dzb, \hspace{1em} \D_2 = p \Dz - q \Dzb,
    \label{eq:mix:eigen}
  \end{equation}
  where $p$ and $q$ are complex scalars, $|p|^2 + |q|^2 = 1$.
  Letting $m_{1,2}$ and $\Gamma_{1,2}$ represent respectively the masses and
  widths of the mass eigenstates $\D_{1,2}$, mixing is usually parameterized by
  the real quantities $x$ and $y$:
  \begin{equation}
    x \equiv \frac{m_1 - m_2}{\Gamma}, \hspace{1em} y \equiv \frac{\Gamma_1 - \Gamma_2}{2 \Gamma}
    \label{eq:mix:xy}
  \end{equation}
  where $\Gamma \equiv \frac{1}{2}\left(\Gamma_1 + \Gamma_2 \right)$.
  \CP is violated in the mixing if $\left|\frac{q}{p}\right| \ne 1$.
  \CPV in the interference between mixing and direct decay is
  parameterized by a real phase $\phi$, which is zero if \CP is conserved.
  The relative argument of $q$ and $p$ is conventionally chosen equal to
  this phase, $\arg\frac{q}{p} = \phi$.

  \LHCb is working towards its first measurements of \CPV and mixing in
  \Dz-\Dzb with lifetime ratios of \mbox{$\Dz \rightarrow \Km \pip$} and
  \mbox{$\Dz \rightarrow \Km \Kp$} decays.
  The lifetime of decays to the \CP-even eigenstate $\Km\Kp$, $\tau(\Km\Kp)$,
  is related to the lifetime of the flavor-specific final state $\Km\pip$,
  $\tau(\Km\pip)$, by the mixing parameters:
  \begin{equation}
    \yCP \equiv \frac{\tau(\Km\pip)}{\tau(\Km\Kp)} - 1 = y \cos\phi 
        - \frac{1}{2}\left(\left|\frac{q}{p}\right| - \left|\frac{p}{q}\right|\right) x \sin\phi.
    \label{eq:mix:ycp}
  \end{equation}
  If \CP is conserved, $\yCP = y$.
  The asymmetry in the lifetimes of \Dz and \Dzb decays to the \CP eigenstate
  $\Km\Kp$ is related to the \CPV and mixing parameters by
  \begin{equation}
    \AGamma \equiv \frac{\tau(\Dzb \to \Km\Kp) - \tau(\Dz \to \Km\Kp)}{\tau(\Dzb \to \Km\Kp) + \tau(\Dz \to \Km\Kp)}
        = \frac{1}{2}\left(\left|\frac{q}{p}\right| - \left|\frac{p}{q}\right|\right) y \cos\phi - x \sin\phi.
    \label{eq:mix:Agamma}
  \end{equation}
  \Dstar-tagged candidates are used in the measurement of \AGamma, while \yCP
  can be measured with the larger untagged sample.

  \begin{figure}[hpb]

    \centerline{
      \subfloat[\Dz]{\label{fig:mix:deltam:Dz}\psfig{file=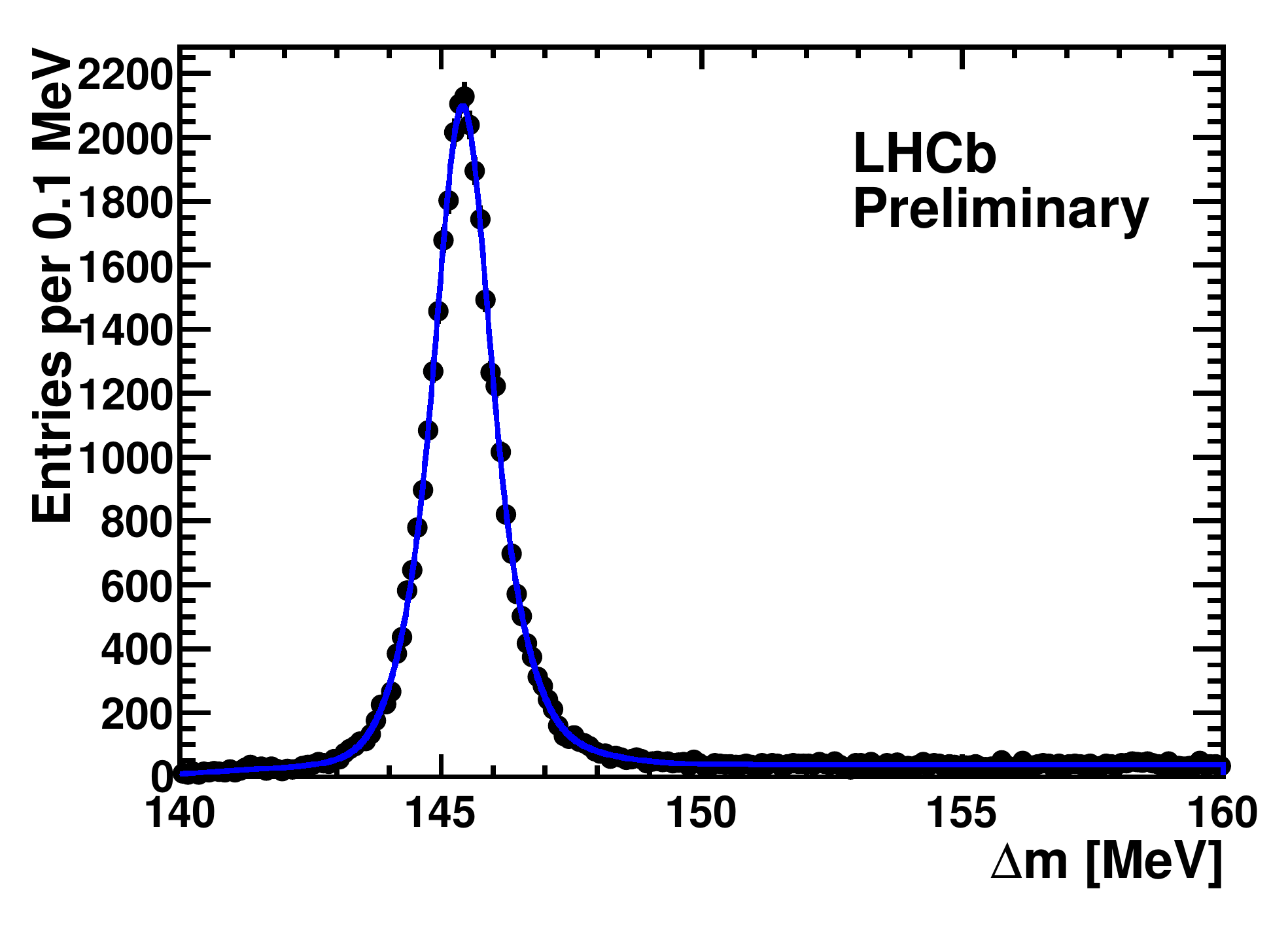,width=2.49in}}%
      \subfloat[\Dzb]{\label{fig:mix:deltam:Dzb}\psfig{file=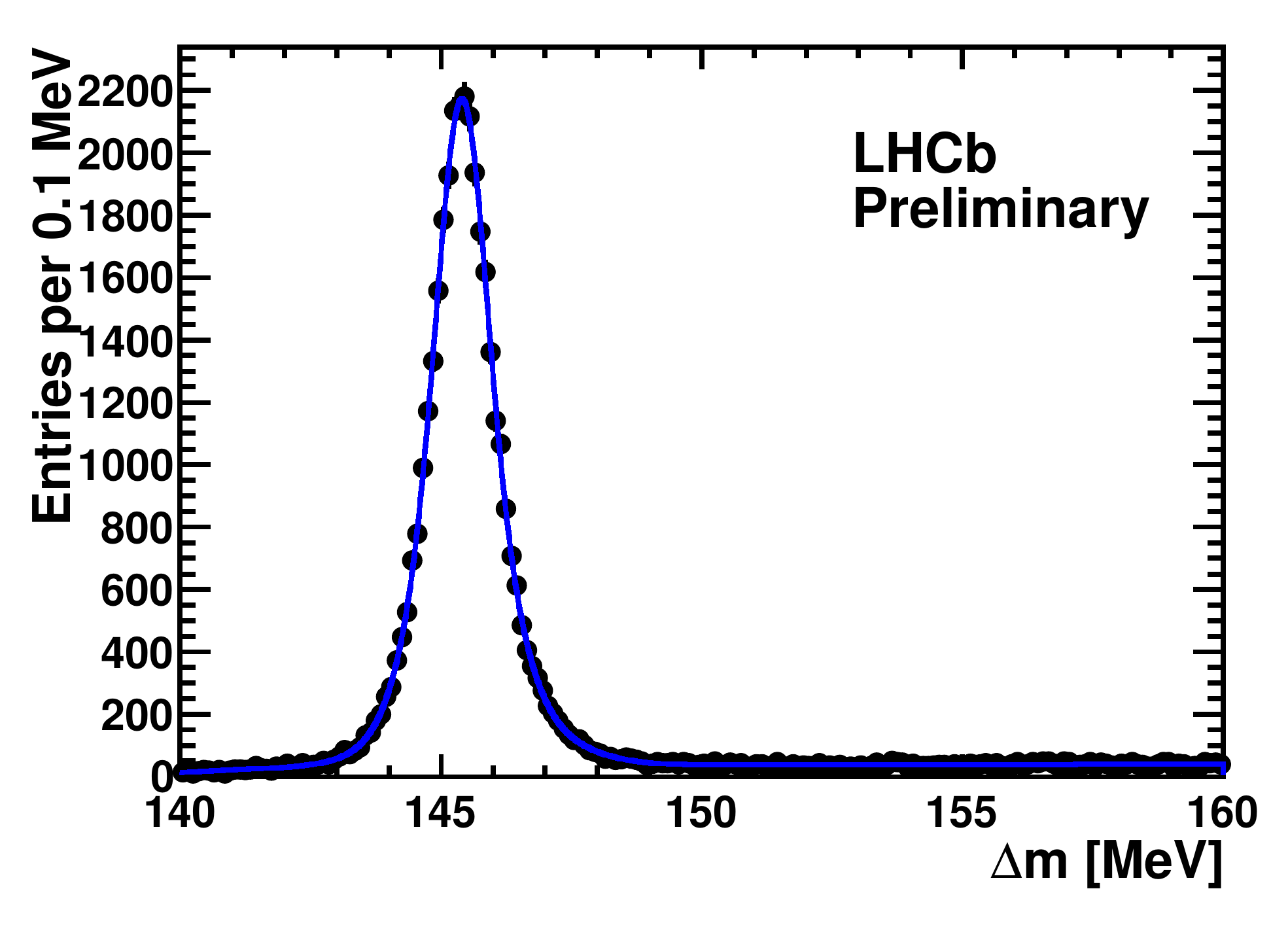,width=2.49in}}}

    \vspace*{8pt}
    \caption{Distributions of the mass difference, $\Delta m$, between
      reconstructed \Dz(\Dzb) candidates and their reconstructed parent
      \Dstarp(\Dstarm) candidates for decays $\Dstarp \rightarrow \Dz \pip$,
      $\Dz \rightarrow \Km \pip$ (c.c.).
     \label{fig:mix:deltam}}
  \end{figure}

  In the 2010 run, we collected a sample of untagged
  \mbox{$\Dz \to \Km \Kp$} decays comparable in size to those of recent
  Belle and BaBar measurements.\cite{Staric:2007dt,Auber:2009ck}
  In 2011-2012, \LHCb expects to have the world's largest charm sample in this
  mode.
  The measurements of \yCP and \AGamma are currently blinded.
  As a test, the \AGamma analysis was applied to a subset of the
  2010 data in the right-sign (RS) control channel $\Dz \rightarrow \Km \pip$.
  Figure~\ref{fig:mix:deltam} shows the distributions of the differences
  \Delm between the masses of the reconstructed \Dz candidates and their
  parent \Dstarp candidates for the RS validation sample.
  The purity of the sample is better than $90\%$.

  The trigger and selection criteria necessary in \lhc collisions introduce a
  proper-time acceptance for the reconstructed \Dz decays.
  Levels of combinatoric backgrounds are large near the primary interaction
  vertex (PV).
  The most powerful signal/background discriminants exploit the relatively long
  lifetime of \D mesons, requiring some signature of separation between the
  PV and reconstructed \D.
  Unbiased time-dependent measurements require careful treatment of the
  acceptance effects of these discriminants.
  We are pursuing two strategies.
  The first is to partition the data set into bins of proper time.
  The observables \yCP and \AGamma can be extracted from the ratios of two
  proper-time distributions, and hence from the distribution of the ratios of
  yields in proper-time bins.
  When the two decay distributions involved have the same final state, as in
  the case of \AGamma, almost all acceptance effects cancel.
  The second method is the event-by-event evaluation of the proper-time
  acceptance by the swimming method.\cite{Gersabeck:1217589,Aaltonen:2010ta}
  For each selected candidate, a single-event acceptance function is calculated
  by determining whether or not the candidate would have passed the selections
  had it decayed at a different proper time.
  This method can be applied exactly to \LHCb data with the original trigger
  and selection software.
  The acceptance thus evaluated is incorporated into an unbinned fit to the
  proper time distribution to measure the lifetime.

  \begin{figure}[hpbt]

    \centerline{
      \subfloat[\Dz]{\label{fig:mix:t:Dz}\psfig{file=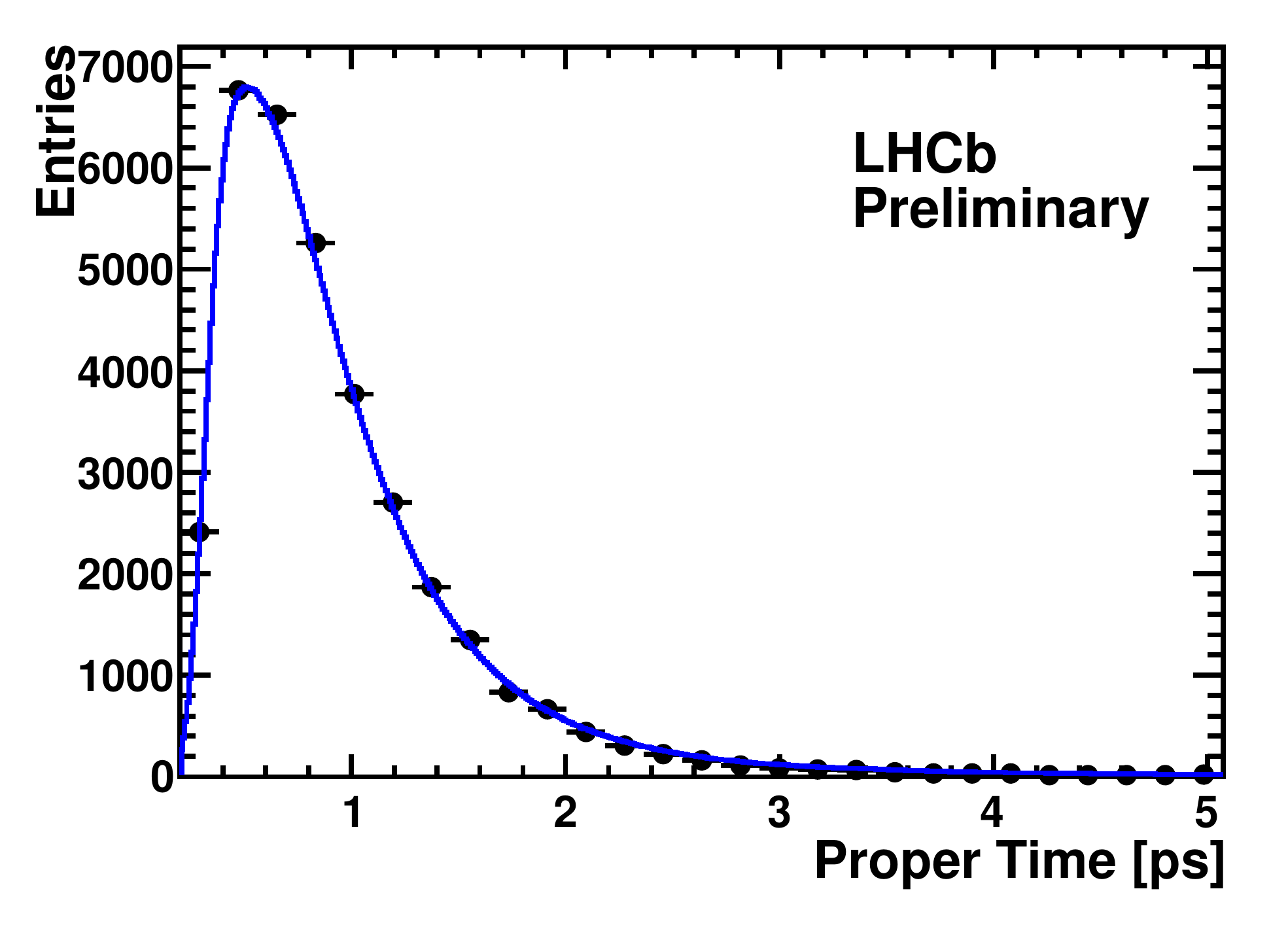,width=2.49in}}%
      \subfloat[\Dzb]{\label{fig:mix:t:Dzb}\psfig{file=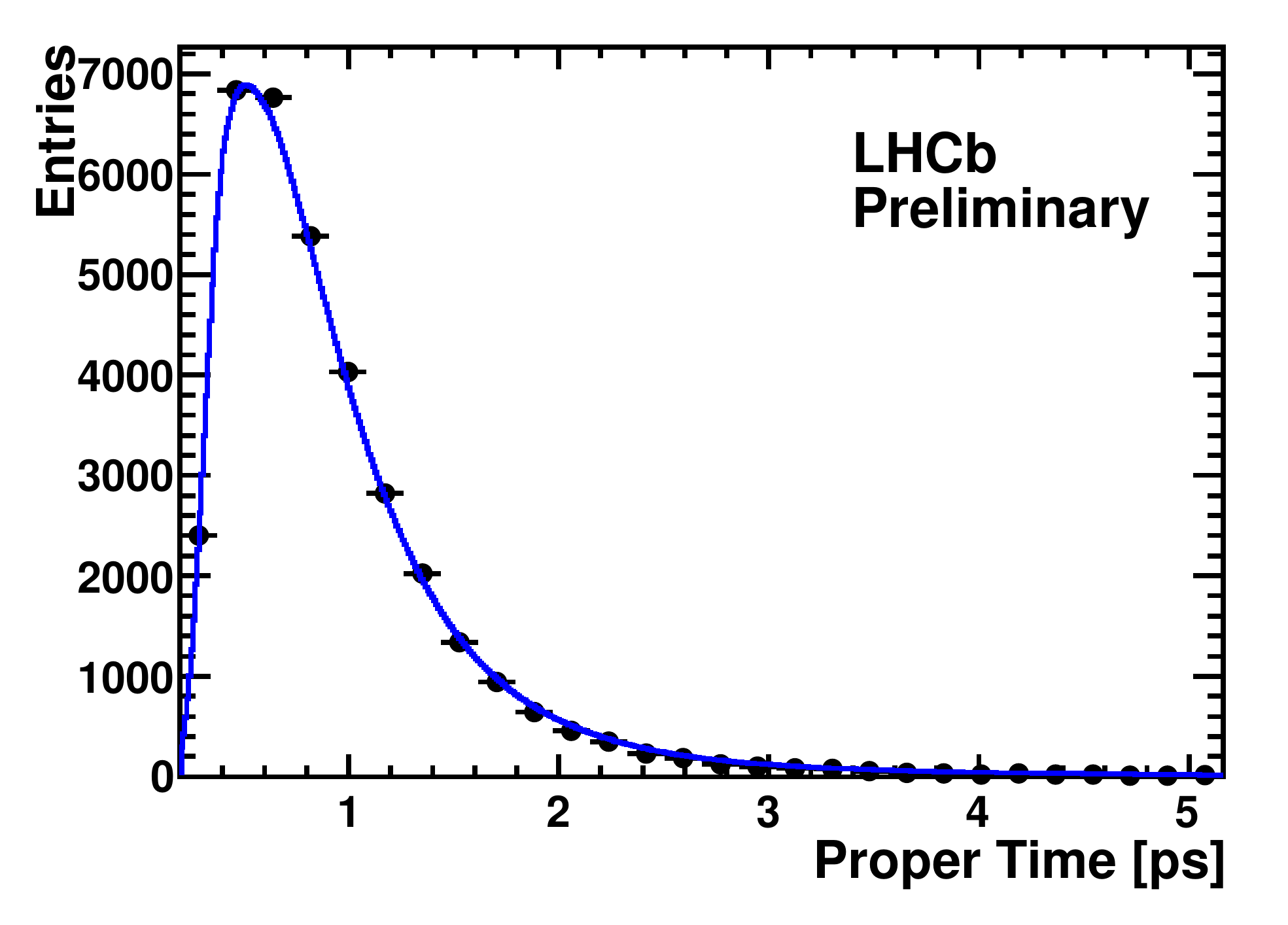,width=2.49in}}}

    \vspace*{8pt}
    \caption{Distributions of the reconstructed proper time of \Dz(\Dzb)
      candidates for decays $\Dstarp \rightarrow \Dz \pip$,
      $\Dz \rightarrow \Km \pip$ (c.c.).
      The line on each plot is the result of a likelihood fit incorporating
      per-event acceptance distributions computed with the swimming method.
     \label{fig:mix:t}}
  \end{figure}

  Another key component of time-dependent charm analysis is the separation
  of \Dz mesons produced at the PV (prompt) from those produced in the decays of
  $b$-hadrons (secondary).
  Because $b$-hadrons also fly before decaying, the apparent proper times of
  secondary \Dz will have a large positive bias.
  Hence, it is vital to statistically separate the two.
  The impact parameter (IP) $\chi^2$ of the \Dz is a powerful discriminant.
  In the binned lifetime measurement, a fit to the IP $\chi^2$ is part of
  the yield measurement in each bin.
  In the unbinned method with event-by-event acceptance, the IP $\chi^2$
  distribution is incorporated into a multi-dimensional likelihood fit.
  Figure~\ref{fig:mix:t} shows the proper-time distributions for the tagged
  RS validation sample.
  The line on the plot is the result of the unbinned multi-dimensional
  likelihood fit.

\section{Summary}
\label{sec:sum}

  \LHCb had a successful year of data taking in 2010, collecting \lumiyr of
  $\proton\proton$ collisions at $\sqrt{s} = 7\,\tev$.
  Charm hadron decays were recorded with high efficiency and in large
  quantities in many channels.
  We have produced our first precision charm \CPV measurement with this
  data:  the difference between the time-integrated \CP asymmetries of
  $\Dz \rightarrow \Km\Kp$ and $\Dz \rightarrow \pim\pip$ decays is measured
  to be $\Delta\ACP = \left(-0.28 \pm 0.70 \pm 0.25 \right)\%$.
  A broad program of charm \CPV and mixing measurements is underway and
  further results in more channels are soon to follow.
  The strategies for controlling key systematic effects are mature,
  and the statistical precision possible with the \LHCb data set is already
  approaching a level comparable with those of the \B-factories in key
  measurements.
  With the large data set expected in 2011-2012, \LHCb is poised to become a
  leader in charm physics.



\begin{thebibliography}{10}

\bibitem{Alves:2008zz}
{\bf LHCb} Collaboration, A.~A. Alves {\em et al.}
  \href{http://dx.doi.org/10.1088/1748-0221/3/08/S08005}{{\em JINST} {\bf 3}
  (2008)  S08005}.

\bibitem{LHCb-CONF-2010-013}
{\bf LHCb} Collaboration, {Conference Report}   LHCb-CONF-2010-013, Dec, 2010.

\bibitem{Bianco:2003vb}
S.~Bianco, F.~L. Fabbri, D.~Benson, and I.~Bigi {\em Riv. Nuovo Cim.} {\bf
  26N7} (2003)  1--200.

\bibitem{Grossman:2006jg}
Y.~Grossman, A.~L. Kagan, and Y.~Nir
  \href{http://dx.doi.org/10.1103/PhysRevD.75.036008}{{\em Phys. Rev.} {\bf
  D75} (2007)  036008}.

\bibitem{LHCb-CONF-2011-023}
{\bf LHCb} Collaboration, {Conference Report}   LHCb-CONF-2011-023, May, 2011.

\bibitem{Aubert:2007if}
{\bf BaBar} Collaboration, B.~Aubert {\em et al.}
  \href{http://dx.doi.org/10.1103/PhysRevLett.100.061803}{{\em Phys. Rev.
  Lett.} {\bf 100} (2008)  061803}.

\bibitem{Staric:2008rx}
{\bf Belle} Collaboration, M.~Staric {\em et al.}
  \href{http://dx.doi.org/10.1016/j.physletb.2008.10.052}{{\em Phys. Lett.}
  {\bf B670} (2008)  190--195}.

\bibitem{CDF-10296}
{\bf CDF} Collaboration, {CDF note}   {10296}, Feb, 2011.

\bibitem{Nakamura:2010zzi:D0mix}
{\bf Particle Data Group} Collaboration, K.~Nakamura {\em et al.},
  \href{http://dx.doi.org/10.1088/0954-3899/37/7A/075021}{``{$D^0$-$\overline{D}^0$
  Mixing},''} in {\em {Review of particle physics}}, vol.~G37, p.~075021.
\newblock
2010.
\newblock

\bibitem{Staric:2007dt}
{\bf Belle} Collaboration, M.~Staric {\em et al.}
  \href{http://dx.doi.org/10.1103/PhysRevLett.98.211803}{{\em Phys. Rev. Lett.}
  {\bf 98} (2007)  211803}.

\bibitem{Auber:2009ck}
{\bf BaBar} Collaboration, B.~Aubert {\em et al.}
  \href{http://dx.doi.org/10.1103/PhysRevD.80.071103}{{\em Phys. Rev.} {\bf
  D80} (2009)  071103}.

\bibitem{Gersabeck:1217589}
M.~Gersabeck, V.~V. Gligorov, J.~Imong, and J.~Rademacker,   LHCb Public note
  LHCb-PUB-2009-022, Nov, 2009.

\bibitem{Aaltonen:2010ta}
{\bf CDF} Collaboration, T.~Aaltonen {\em et al.}
  \href{http://dx.doi.org/10.1103/PhysRevD.83.032008}{{\em Phys. Rev.} {\bf
  D83} (2011)  032008}.

\end{thebibliography}

\providecommand{\href}[2]{#2}
\begingroup\raggedright\endgroup

\end{document}